\documentclass[twocolumn]{emulateapj}
\usepackage{graphics,graphicx}
\usepackage{color}
\usepackage{natbib}

\slugcomment{Accepted to the Astrophysical Journal}
\shorttitle{Circinus Galaxy in $\gamma$ rays}
\shortauthors{Hayashida et al.}

\begin{document}

\title{Discovery of GeV Emission from the Circinus galaxy with the \textit{Fermi}-LAT}

\author{Masaaki~Hayashida\altaffilmark{1,\,2,\,3}, 
{\L}ukasz~Stawarz\altaffilmark{4,\,5},
Chi~C.~Cheung\altaffilmark{6}, 
Keith~Bechtol\altaffilmark{7}, 
Greg~M.~Madejski\altaffilmark{2}, 
Marco~Ajello\altaffilmark{8}, 
Francesco~Massaro\altaffilmark{2},
Igor~V.~Moskalenko\altaffilmark{2}, 
Andrew~Strong\altaffilmark{9}, 
and Luigi~Tibaldo\altaffilmark{2}
}

\altaffiltext{1}{Institute for Cosmic Ray Research, University of Tokyo, 5-1-5 Kashiwanoha, Kashiwa, Chiba, 277-8582, Japan}
\altaffiltext{2}{Kavli Institute for Particle Astrophysics and Cosmology, SLAC National Accelerator Laboratory,  Stanford University, 2575 Sand Hill Road M/S 29, Menlo Park, CA 94025, USA}
\altaffiltext{3}{email: mahaya@icrr.u-tokyo.ac.jp}
\altaffiltext{4}{Institute of Space and Astronautical Science, JAXA, 3-1-1 Yoshinodai, Chuo-ku, Sagamihara, Kanagawa 252-5210, Japan}
\altaffiltext{5}{Astronomical Observatory, Jagiellonian University, ul. Orla 171, 30-244 Krak\'ow, Poland}
\altaffiltext{6}{Space Science Division, Naval Research Laboratory, Washington, DC 20375-5352, USA}
\altaffiltext{7}{Kavli Institute for Cosmological Physics, University of Chicago, 5640 South Ellis Avenue, Chicago, IL 60637}
\altaffiltext{8}{Space Sciences Laboratory, 7 Gauss Way, University of California, Berkeley, CA 94720-7450, USA}
\altaffiltext{9}{Max-Planck Institut f\"ur extraterrestrische Physik, 85748 Garching, Germany}

\label{firstpage}

\begin{abstract}

We report the discovery of $\gamma$-ray emission from the Circinus 
galaxy using the Large Area Telescope (LAT) onboard the \textit{Fermi
  Gamma-ray Space Telescope}. Circinus is a nearby ($\sim4$ Mpc)
starburst with a heavily obscured Seyfert-type active nucleus, bipolar
radio lobes perpendicular to the spiral disk, and kpc-scale 
jet-like structures. Our analysis of 0.1--100 GeV events collected
during 4 years of LAT observations reveals a significant ($\simeq
7.3\,\sigma$) excess above the background. We find no indications of variability or spatial
extension beyond the LAT point-spread function. A power-law model used to describe 
the $0.1-100$\,GeV $\gamma$-ray spectrum yields a flux of 
$(18.8\pm5.8)\times10^{-9}$\,ph\,cm$^{-2}$\,s$^{-1}$ and photon index
$2.19\pm0.12$, corresponding to an isotropic $\gamma$-ray 
luminosity of $3 \times 10^{40}$\,erg\,s$^{-1}$. 
This observed $\gamma$-ray luminosity exceeds the luminosity expected from cosmic-ray
interactions in the interstellar medium and inverse Compton radiation from the radio
lobes. Thus the origin of the GeV excess requires further investigation.

\end{abstract}


\section{Introduction}
\label{sec:intro}

Active galactic nuclei (AGN) often possess fast nuclear outflows and collimated
radio-emitting jets powered by accretion onto supermassive black holes 
\citep[e.g.,][]{Krolik99}. Relativistic jets produced in radio-loud AGN (such as
blazars and radio galaxies), which are typically hosted by early-type galaxies,
are well-established sources of Doppler-boosted $\gamma$-ray emission, 
dominating the extragalactic source population in the GeV range \citep{2FGL}. 
The bulk of the observed $\gamma$-ray jet emission in such systems
is believed to originate within a few parsecs of the central black hole \cite[e.g.,][]{3C279}. 
However, \textit{Fermi}-LAT observations of nearby radio galaxies indicate that 
large-scale structures --- hereafter `lobes' or `bubbles' --- formed by jets outside their 
hosts due to interactions with the intergalactic medium, can also be bright GeV 
emitters (\citealt{CenA}; see also \citealt{Katsuta13}).

Radio-quiet AGN, the most luminous of which are classified as Seyferts by their 
optical spectra, are typically hosted by late-type galaxies.
The `radio-quiet' label does not necessarily mean these sources are 
completely ‘radio-silent', and the nuclear jets 
found in these systems are non-relativistic 
and comparatively low-powered \citep[e.g.,][]{Lal11}, so that AGN and starburst
activities may both be equally important factors in forming of kpc-scale outflows and lobes
\citep[e.g.,][]{Gallimore06}. Seyfert galaxies seem $\gamma$-ray quiet as a class~\citep{Teng11,Seyferts}. 

Besides radio-loud AGN, another class of established extragalactic 
$\gamma$-ray sources consists of galaxies lacking any pronounced nuclear activity, but
experiencing a burst of vigorous star formation. The observed GeV
emission from these sources is most readily explained by interactions of galactic 
cosmic rays (CRs) with ambient matter and radiation fields of the interstellar medium 
\citep[ISM;][and references therein]{Starforming}. Starburst-driven outflows routinely found in such systems may lead to 
the formation of large-scale bipolar structures extending perpendicular to the galactic disks, 
and somewhat resembling jet-driven AGN lobes \citep{Veilleux05}. 

Circinus is one of the nearest \citep[distance $D = 4.2 \pm
 0.7$\,Mpc;][]{Tully09} and most extensively studied composite
starburst/AGN systems, but its location
behind the intense foreground of the Milky Way disk, ($l$, $b$) =
($311.3^{\circ}$, $-3.8^{\circ}$), resulted in exclusion
from several previous population studies using LAT
data~\citep{Seyferts, Starforming}. Circinus exhibits a polarized broad H$\alpha$ line
 with FWHM $\sim3000$\,km\,s$^{-1}$ \citep{Oliva98}, evidencing a
 heavily obscured Seyfert 2 nucleus which ranks as
 the third brightest Compton-thick AGN \citep{Yang09}. Radio
 observations have mapped bipolar bubbles extending orthogonal to the
 spiral host, as well as kpc-scale jet-like structures likely 
responsible for the formation of the lobes \citep{Elmouttie98}. The extended morphology 
of the system including the lobes has been recently resolved and
studied in X-rays \citep{Mingo12}.

\begin{figure*}[!t]
\begin{center}
\includegraphics[width=10cm]{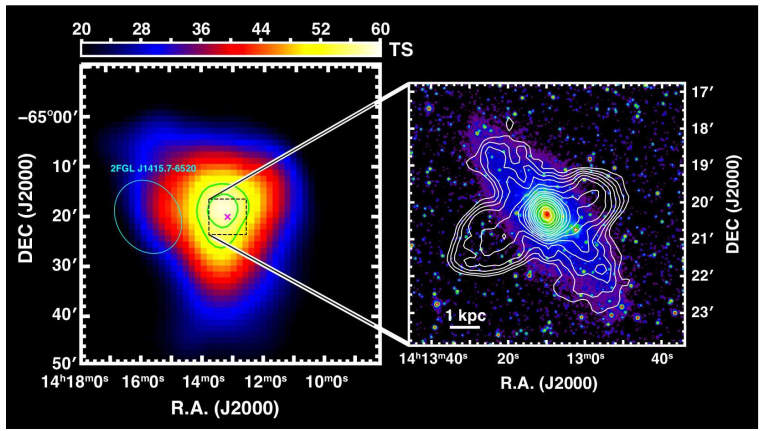}
\caption[]{The left color map represents the 
$1^{\circ} \times 1^{\circ}$ ($0^{\circ}.02$ pixels)
 spatial TS variation of the $>100$\,MeV $\gamma$-ray
  signal excess centered on
  Circinus. The green lines denote positional errors of the
  $\gamma$-ray excess at 68\,\% and 95\,\% confidence levels for inner
  and outer lines, respectively. The magenta cross indicates the
  position of the galaxy core while the cyan ellipse
  corresponds to the positional error at 95\% confidence of
  2FGL\,J1415.7$-$6520. The black square corresponds to the area of the
  right panel, 
  which shows the Australia Telescope Compact Array (ATCA) 
  1.4 GHz radio contours at $\simeq 20''$
  resolution \citep{Elmouttie98} 
  superposed with the 2MASS $H$-band color image~\citep{2MASS}; 
  here the galactic disk extends in the
  NE--SW direction, while the radio lobes in the SE--NW direction.\label{fig:LAT}}
\end{center}
\end{figure*}

Here, we report the discovery of a significant $\gamma$-ray
excess positionally coinciding with the Circinus galaxy
using 4 years of \textit{Fermi}-LAT data and discuss possible origins for this emission.

\section{LAT Data Analysis and Results}
\label{sec:LAT}

The LAT is a pair-production telescope onboard the \textit{Fermi} satellite with large effective 
area (6500\,cm$^2$ on axis for $>1$\,GeV photons) and large field of view (2.4\,sr), 
sensitive from $20$\,MeV to $> 300$\,GeV 
\citep{LAT}. Here, we analyzed LAT data for the Circinus region collected 
between 2008 August 5 and 2012 August 5, following the standard 
procedure\footnote{\texttt{http://fermi.gsfc.nasa.gov/ssc/data/analysis/}}, 
using the LAT analysis software \texttt{ScienceTools} \texttt{v9r29v0} with the 
\texttt{P7SOURCE\_V6} instrument response functions. Events in the energy 
range 0.1--100\,GeV were extracted within a $17^{\circ} \times 17^{\circ}$ 
region of interest (RoI) centered on the galaxy core position (RA\,$= 213{\fdg}2913$, 
Dec\,$=-65{\fdg}3390$: J2000). $\gamma$-ray fluxes and spectra 
were determined by a maximum likelihood fit with \texttt{gtlike}
for events divided into $0{\fdg}1$-sized pixels and 30 uniformly-spaced 
log-energy bins. 
The background model included all known $\gamma$-ray sources within the RoI 
from the 2nd LAT catalog \citep[2FGL:][]{2FGL}, except for 2FGL\,J1415.7$-$6520, 
located $0{\fdg}266$ away from Circinus. Additionally, the model included 
the isotropic and Galactic diffuse emission components\footnote{\texttt{iso\_p7v6source.txt} 
and \texttt{gal\_2yearp7v6\_v0.fits}}; flux normalizations for the diffuse 
and background sources were left free in the fitting procedure.

Our analysis yields a test statistic (TS) of 58 for the maximum likelihood 
fit when placing a new candidate source at the core position of Circinus,
corresponding to a formal detection significance of 
$\simeq 7.3\,\sigma$. Figure~\ref{fig:LAT} shows the spatial variation
in the TS value for the candidate source when evaluated over a grid of
positions around the galaxy (TS map). 
Circinus is located within the 68\% confidence region for the
direction of the $\gamma$-ray excess. The distribution of the $\gamma$-ray excess is consistent 
with a point-like feature, but we note that the LAT does not have sufficient 
angular resolution~\citep[see e.g.,][]{PSF} to spatially resolve the core, the radio lobes, or the galactic disk 
(all shown in the right panel of Figure~\ref{fig:LAT}) of the system.

\begin{figure}[!t]
\begin{center}
\includegraphics[width=\columnwidth]{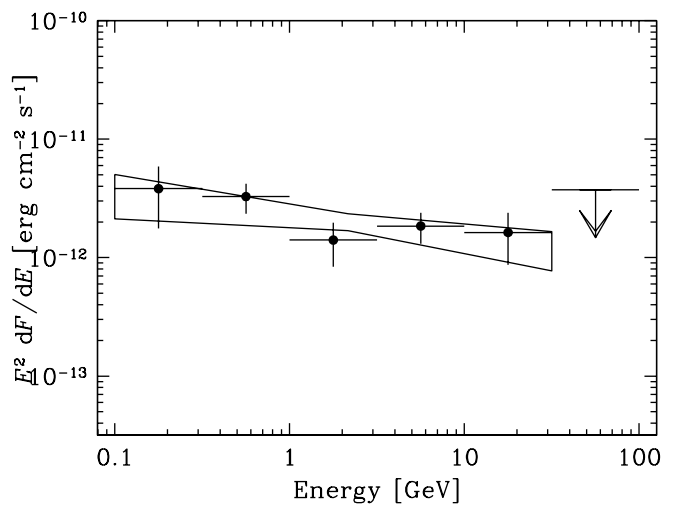}
\caption[]{GeV spectrum of Circinus as measured by
  LAT. The black lines represent the best-fit PL model
  ($1\,\sigma$ confidence band). The vertical bars denote $1\,\sigma$ statistical errors for the flux estimates within energy bins given by horizontal bars. The arrow denotes a 95\% confidence level upper limit. \label{fig:spec}}
\end{center}
\end{figure}

Assuming 2FGL\,J1415.7$-$6520 is an additional $\gamma$-ray source distinct from 
the Circinus galaxy, we repeat the analysis including 2FGL\,J1415.7$-$6520 in the background model. 
The $\gamma$-ray excess at the position of Circinus yields in such a case a TS of 52, while no 
significant excess can be seen at the formal position of 2FGL\,J1415.7$-$6520 (TS\,$\simeq$\,5.6). 
Our analysis confirms therefore a single point-like $\gamma$-ray excess positionally coinciding with
Circinus, which may be identified
with 2FGL\,J1415.7$-$6520 after the localization of the 2FGL source is refined 
using the 4\,year-accumulation of LAT data.
The 2FGL analysis flagged 2FGL~J1415.7$-$6520 as a source with a low signal-to-background ratio (2FGL flag was set to `4’), meaning its position can be relatively strongly affected by systematic uncertainties in the Galactic diffuse emission model~\citep[see also][]{diffuse2}, and the overall positional error of 2FGL~J1415.7$-$6520 could be larger than indicated. Our analysis here of 4 years of accumulated LAT data for the source allowed it to be detected at higher energies, thus decreasing the systematic uncertainty in the location.

\begin{figure}[!t]
\begin{center}
\includegraphics[width=5cm]{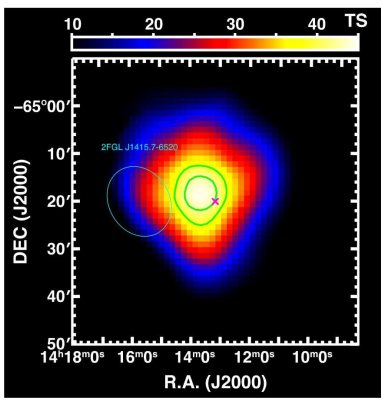}
\caption[]{The color map represents the 
$1^{\circ} \times 1^{\circ}$ ($0^{\circ}.02$ pixels)
 spatial TS variation of the $>1$\,GeV $\gamma$-ray
  signal excess centered on
  Circinus. The green lines denote positional errors of the
  $\gamma$-ray excess at 68\,\% and 95\,\% confidence levels. 
  The magenta cross indicates the
  position of the galaxy core while the cyan ellipse
  corresponds to the positional error at 95\% confidence of
  2FGL\,J1415.7$-$6520. \label{fig:LAThigh}}
\end{center}
\end{figure}

\begin{figure}[!t]
\begin{center}
\includegraphics[width=\columnwidth]{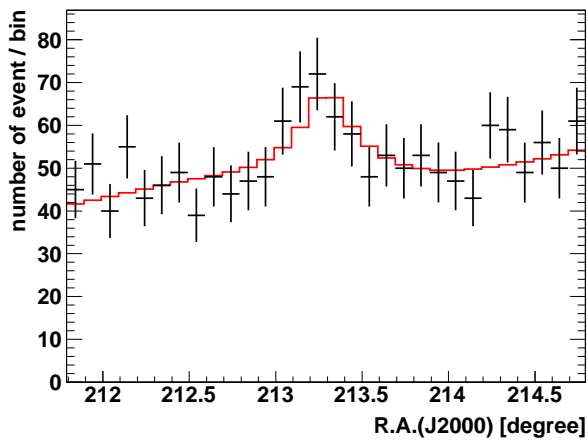}
\caption[]{Observed counts profile of $>1$\,GeV $\gamma$ rays for the Circinus 
source with events within $\pm0.5$ degree 
in Declination (north-south) projected along the R.A. (east-west) direction. 
The red curve represents the best-fit emission model with the point-like source and diffuse emission. 
The observed profile is consistent with a point-like source, 
indicating no significant spatial extension in the $\gamma$-ray excess associated with Circinus.
 \label{fig:proj}}
\end{center}
\end{figure}

A simple power-law model (PL) adequately describes the $\gamma$-ray spectrum of the 
source, yielding a $\gamma$-ray flux $\mathcal{F}_{\rm >0.1\,GeV} = (18.8\pm5.8) 
\times10^{-9}$\,ph\,cm$^{-2}$\,s$^{-1}$ and photon index $\Gamma =
2.19\pm0.12$, corresponding to an isotropic $0.1-100$\,GeV luminosity of 
$L_{\gamma} \simeq (2.9\pm0.5) \times 10^{40}$\,erg\,s$^{-1}$. The best-fit model together 
with flux points of the spectral energy distribution (SED) are plotted
in Figure~\ref{fig:spec}. We also perform 
spectral fits using a broken power-law model and a log-parabola model, 
but no significant improvement ($< 1.2\,\sigma$) can be seen in the likelihood values when compared 
with the PL parametrization. The results of the spectral fits 
for Circinus are summarized in Table~\ref{tab:fit}.
For comparison, we also derived $\gamma$-ray fluxes and spectra for four LAT-detected starburst 
galaxies~\citep{Starforming}
with 4-yr LAT data by performing the same analysis procedure described above,
and summarize the results in Table~\ref{tab:fit}.

Since Circinus is located near the Galactic plane, it is important 
to consider systematic errors caused by uncertainties in the Galactic 
diffuse emission model. We test eight alternative diffuse models
\citep[following][]{SNR,dePal13},
and find that the derived source flux above 0.1\,GeV shifts by no more
than 3\%, which is much smaller than the statistical uncertainty.
The detection significance at the source position remains stable for 
the different alternative diffuse models (TS\,$\simeq 58-60$).

We also analyze a limited event sample with an energy range
$1-100$\,GeV within a $10^{\circ} \times 
10^{\circ}$ RoI. In this energy range, reduced systematic effects related to diffuse
emission modeling are expected, owing in part to the improved spatial
resolution of the LAT. The power-law character of the $\gamma$-ray 
continuum is confirmed with $\Gamma = 2.19\pm0.41$, and we find that
much of the statistical power of our search comes from photons above 1 GeV (TS\,$=$\,42).
The extrapolated flux above 0.1\,GeV from the higher energy analysis corresponds to 
$(16.4\pm5.6)\times10^{-9}$\,ph\,cm$^{-2}$\,s$^{-1}$, in good agreement 
with the result obtained using events from 0.1\,GeV.
Figure~\ref{fig:LAThigh} shows a TS map created using events 
with this energy band, and we have confirmed that Circinus is located 
at the edge of the 68\% confidence contour of the $\gamma$-ray source. 
We also produced an observed counts profile of Circinus and compared 
it with the best-fit point-like source model as shown in Figure~\ref{fig:proj}. 
The observed counts profile is consistent with the model shape of a point-like source, 
indicating no significant spatial extension of the $\gamma$-ray excess.

\begin{figure}[!t]
\begin{center}
\includegraphics[width=\columnwidth]{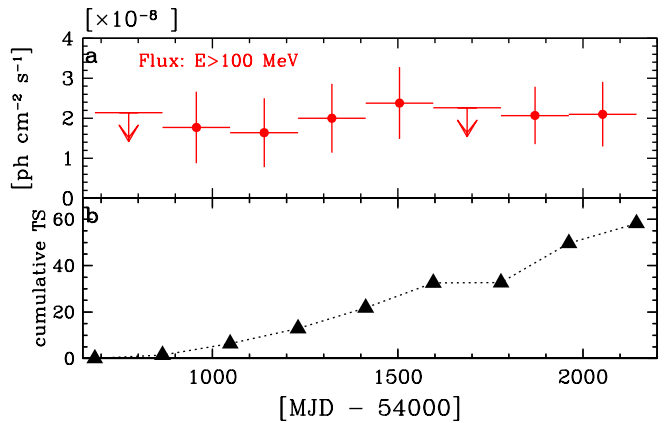}
\caption[]{The top panel (a) shows the light curve of Circinus above 100\,MeV, 
binned in half-year intervals. For bins with TS $<$ 4, a 95\,\% confidence 
level upper limit is plotted. The bottom panel (b) presents the cumulative 
TS of the $\gamma$-ray excess at the source position. \label{fig:LC}}
\end{center}
\end{figure}

\begin{table*}[!th]
{\scriptsize
\caption{Starburst galaxies detected with \textit{Fermi}-LAT. \label{tab:fit}}
\begin{center}
\begin{tabular}{cccccccccc}
\hline
Name & $D$ & $L_{\rm IR}/10^{44}$& $L_{\rm R}/10^{38}$ & TS  & $\mathcal{F}_{\rm >0.1\,GeV}$ & $\Gamma$ & $L_{\gamma}/10^{40}$ &  $L_{\gamma}/L_{\rm IR}$  & $L_{\gamma}/L_{\rm R}$ \\
 & Mpc & erg\,s$^{-1}$ & erg\,s$^{-1}$ &  & $10^{-9}$ph\,cm$^{-2}$\,s$^{-1}$ 	&	& erg\,s$^{-1}$  & $\times 10^{-4}$ &  $\times 10^{-2}$  
\\
\hline 
\hline
Circinus & 4.2  & 0.60& 0.45 & 58 &$18.8\pm5.8$ & $2.19\pm0.12$ & $2.9\pm0.5$ & $ 4.9\pm0.9$  & $6.6\pm1.1$\\
NGC\,253 &  2.5 & 0.69	 &	0.59 & 135 & $10.7\pm2.1$ & $2.18\pm0.09$ & $0.60\pm0.07$ & 	$0.87\pm0.10$ &  $1.02\pm0.12$ \\
M\,82 &  3.4 & 2.0	& 1.5 & 	205 & $16.2\pm2.3$ & $2.25\pm0.08$  & $1.47\pm0.14$	 & 	$0.74\pm0.07$ &   $0.99\pm0.09$  \\
NGC\,4945 & 3.7 & 	1.0 & 1.5 &  42 &  $7.2\pm2.4$ & $2.05\pm0.13$ & $1.17\pm0.23$ &		$1.13\pm0.22$ &  $0.77\pm0.15$ \\
NGC\,1068 &  16.7 & 11 & 23 &  46 & $7.3\pm2.9$ &  $2.29\pm0.19$ & $15.0\pm2.9$ &		$1.41\pm0.27$ & $0.64\pm0.12$ \\
\hline \\ 
\end{tabular}
\end{center}
}
\end{table*}

\begin{figure*}[!t]
\begin{center}
\plottwo{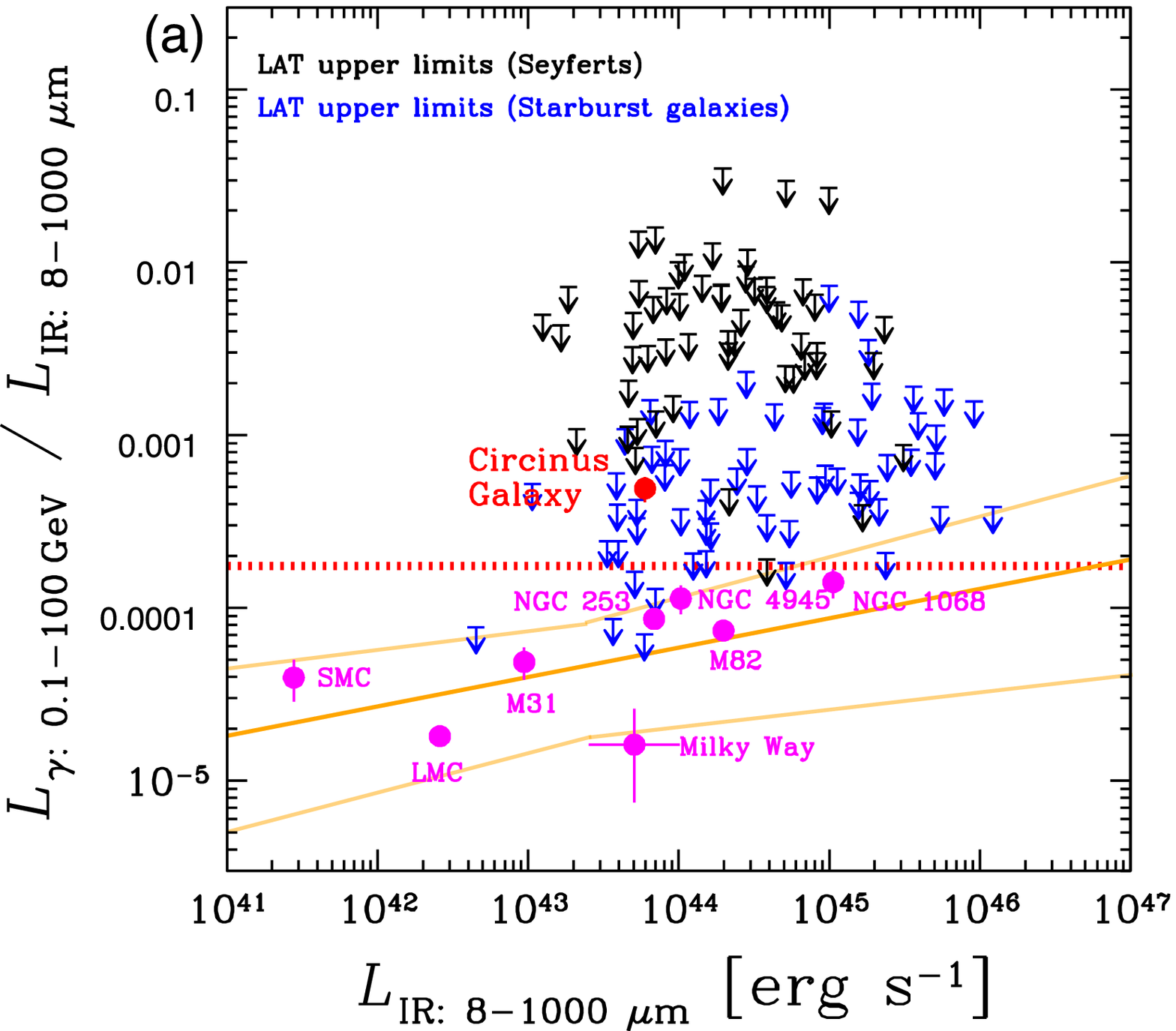}{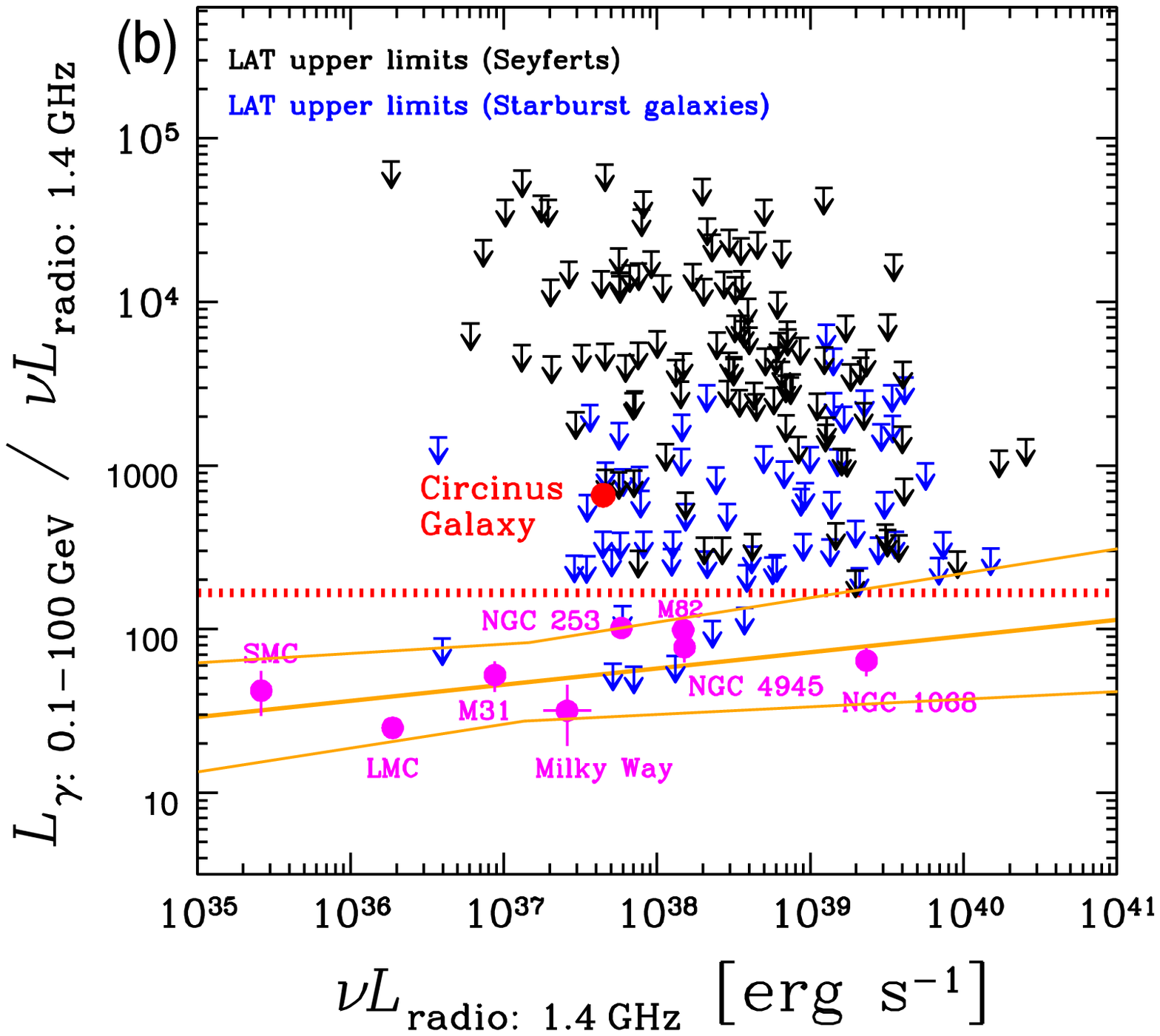}
\caption[]{Comparisons between (a:left) the $\gamma$-ray (0.1-100\,GeV) and
  total IR ($8-1000\,\mu$m), (b:right) the $\gamma$-ray and
  1.4\,GHz radio luminosities of star-forming galaxies and
  Seyferts. Circinus is denoted by the red filled circle. Eight
  $\gamma$-ray detected star-forming galaxies are plotted by magenta
  points. $\gamma$-ray upper limits for non-detected starburst
  galaxies selected from the HCN survey are denoted by blue arrows,
  and for hard X-ray selected Seyfert galaxies by black arrows. The
  orange line represents the best-fit power-law relation between
  $L_{\gamma}$ and $L_{\rm IR}$ (or $\nu L_{\rm radio}$) for star-forming galaxies, and the
  orange bands includes the fit uncertainty and intrinsic dispersion
  around the fitted relation. The dotted red line corresponds to the
  calorimetric limit assuming an average CR luminosity per supernova
  of $\eta E_{\rm SN} = 10^{50}$\,erg. \label{fig:excess}}
\end{center}
\end{figure*}

In order to search for variability in the $\gamma$-ray flux, 
we divided the 4-year observation window into half-year intervals and
derived $\gamma$-ray fluxes above 0.1\,GeV for each time interval using 
a PL model with photon index fixed at the best-fit value 
obtained for the 4-year dataset. No significant variability is found
in the $\gamma$-ray flux ($\chi^2$/d.o.f$\,$=$\,5.2/7$). 
We also checked the cumulative significance of the $\gamma$-ray excess at the 
source position during the 4-year observation; the TS showed a 
gradual increase, as expected for a steady source. Both the source
light curve and the cumulative TS are presented in Figure~\ref{fig:LC}.

\section{Discussion}
\label{sec:discussion}

In Figure~\ref{fig:excess}, we plot the $\gamma$-ray 
luminosity of Circinus, $L_{\gamma}$, normalized by the total IR
($8-1000$\,$\mu$m) luminosity, $L_{\rm IR}$, as a function of $L_{\rm IR}$.  
For comparison, the plot includes $\gamma$-ray upper limits 
(95\% confidence level) derived for a large sample of starbursts 
selected from the HCN survey, 
together with eight $\gamma$-ray detected star-forming galaxies 
(the four Local Group galaxies and four starbursts). 
These all follow \citet{Starforming}, 
with the exception of the four starbursts where the results are from 
our updated 4-yr LAT analysis (Table~\ref{tab:fit}).
In addition, 
we also show the $\gamma$-ray upper limits for the sample of hard X-ray 
selected Seyferts \citep{Seyferts}. The IR 
luminosities are derived from the fluxes measured by the Infrared 
Astronomical Satellite (IRAS) in four bands, following \citet{Sanders96}.

As shown in Figure~\ref{fig:excess}, and discussed quantitatively 
in \citet{Starforming}, the $\gamma$-ray emission related to CRs 
in the ISM scales with the total IR luminosity, which serves as a
proxy for the injected CR power within the supernova remnant
paradigm for galactic CR origin. Circinus is more than five times over-luminous 
in the $\gamma$-ray band when compared with this scaling relation (see also 
Table~\ref{tab:fit}). In fact, the $L_\gamma/L_{\rm IR}$ luminosity 
ratio for Circinus exceeds the so-called `calorimetric limit' 
expected to hold when CRs interact faster than they can escape the
galactic disk, and as much as $\eta \simeq 10\,\%$ energy per supernova 
explosion goes into CR acceleration, $ \eta E_{\rm SN} \simeq 
10^{50}$\,erg \citep[see, e.g.,][]{Lacki11,Starforming}. Circinus also
does not follow the correlation found for star-forming galaxies between 
$L_{\gamma}$ and total monochromatic 1.4\,GHz radio luminosity 
$\nu L_{\rm R}$, by a factor of six (see Table~\ref{tab:fit} and the right panel of Figure~\ref{fig:excess}).
Following the approach of \citet{Starforming}, and assuming
that star-forming galaxies do follow a simple power-law scaling
between $\gamma$-ray and total-IR luminosity, we find that adjusting
either the normalization of this relation or the intrinsic scatter to
accommodate Circinus would tend to overpredict the total number of
galaxies which have been actually detected by LAT.

Let us consider the possibility that the total IR and radio luminosities
listed in Table~\ref{tab:fit} underestimate the star formation rate
(SFR). Whereas standard scaling relations 
using IRAS fluxes yield $\rm{SFR} \simeq 2\,M_{\odot}$\,yr$^{-1}$,
the most recent mid-IR studies of Circinus with \textit{Spitzer}
presented by \citet{For12} suggest $\rm{SFR} \simeq
(3-8)\,M_{\odot}$\,yr$^{-1}$ (depending on the
particular calibration method).
The differences between SFRs derived from mid-IR and total-IR (IRAS)
measurements might be explained if Circinus is a relatively dust-poor system
for which the far-IR luminosity is not a reliable indicator of the SFR.
The SFR converted from the observed total 1.4\,GHz fluxes~\citep{Yun01} is only
$\simeq 1.5\,M_{\odot}$\,yr$^{-1}$, but the radio
luminosity could also under-represent the actual CR injection power if the
interstellar magnetic field strength were below average. 
On the other hand, the total flux density ratio 
$\log \left(F_{\rm 70\,\mu m}/F_{\rm 1.4\,GHz}\right) = 2.40$
demonstrates that Circinus obeys the `far-IR/radio' correlation established for local
star-forming and starburst systems \citep[e.g.,][]{Seymour09}.
The luminosities of NGC\,1068 and NGC\,4945, which are
similar starburst/Seyfert composite galaxies, are consistent with the 
$\gamma$-ray-to-total-IR and $\gamma$-ray-to-radio correlations 
\cite[][see also Figure~\,\ref{fig:excess}]{Starforming}.
These complexities in the Circinus system make a firm conclusion
regarding the attribution of $\gamma$ rays to CR interactions difficult.

Other than CR interactions, the disk coronae or accretion disks of Seyfert galaxies can be 
considered as possible $\gamma$-ray emission sites~\citep[e.g.,][]{Andrzej}. However, LAT 
studies of a large sample of hard X-ray--selected Seyferts devoid of prominent 
jets revealed that such sources are $\gamma$-ray quiet as a class, 
down to the level of $1-10\%$ of the hard X-ray fluxes \citep{Seyferts}. 

One of the striking characteristics of the Circinus system is the presence 
of well-defined radio lobes and kpc-scale jet-like features, the `plumes' 
\citep{Elmouttie98}, which are also resolved at X-ray frequencies 
with \textit{Chandra} \citep{Mingo12}. Both of these might be relevant $\gamma$-ray 
emission sites. To investigate this idea quantitatively, we apply 
standard leptonic synchrotron and inverse-Compton (IC) modeling to 
the radio spectra of the lobes, investigating whether the extrapolation 
of the high-energy emission continuum may account for the GeV 
flux from the system. In the IC calculations we consider seed photons 
provided by the cosmic background radiation, the observed IR--to--optical 
emission of the galactic disk, as well as the UV--to--hard X-ray photon field 
due to the active nucleus, corrected for obscuration \citep[assuming 
the intrinsic UV emission of the accretion disk $\simeq 3 \times 
10^{43}$\,erg\,s$^{-1}$;][]{Prieto10}. 

\begin{figure}[!t]
\begin{center}
\includegraphics[width=\columnwidth]{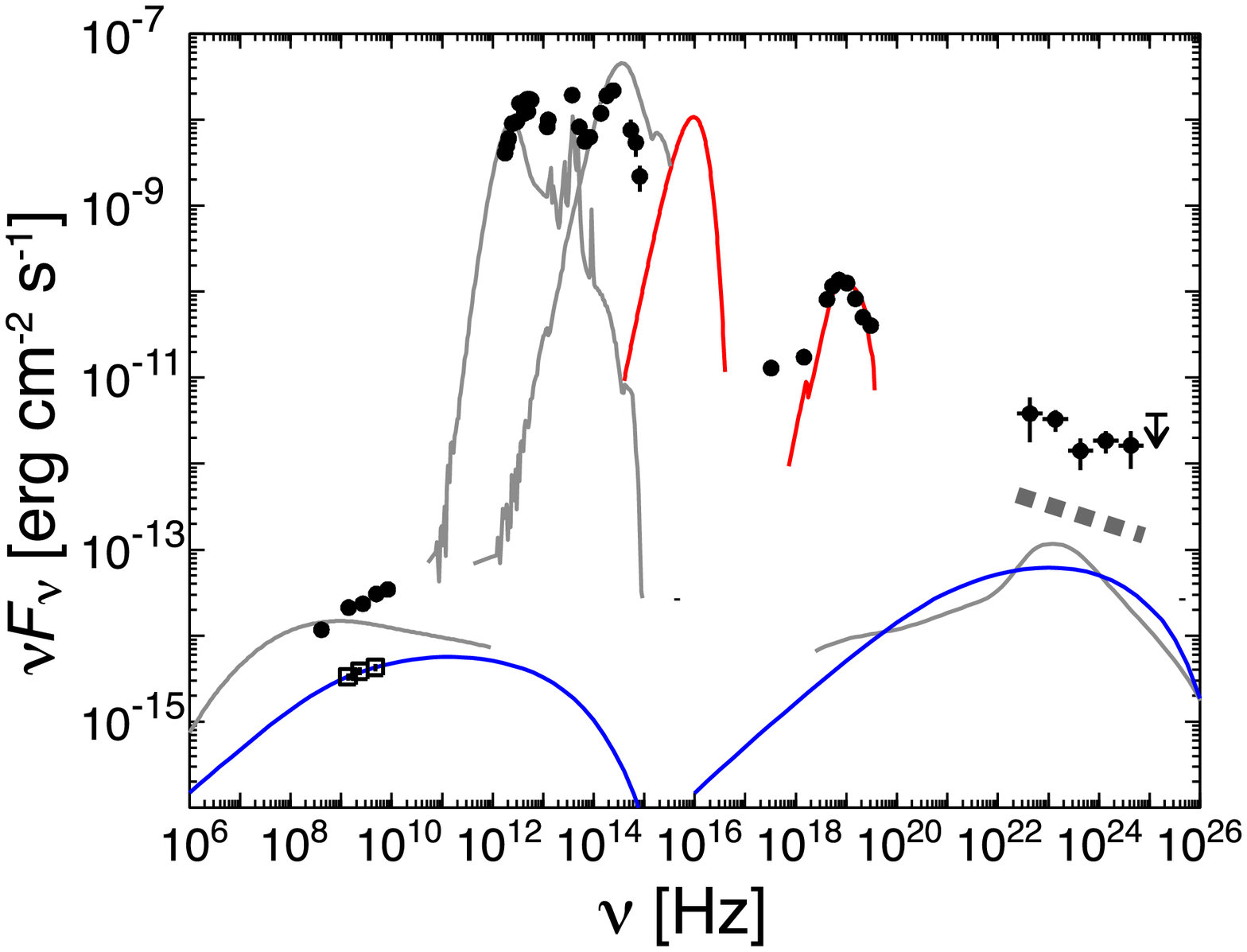}
\caption[]{Broad-band, multi-component SED of Circinus. Black open
  squares represent the total `lobes $+$ plumes' radio fluxes, with
  the assumed $10\%$ uncertainties. Black filled circles denote the
  total (integrated) fluxes of the system from the Parkes Catalog
  \citep[$0.4-8.6$\,GHz;][]{Wright90}, ISO
  \citep[$170-52$\,$\mu$m;][]{Brauher08}, IRAS ($100-25$\,$\mu$m),
  \textit{Spitzer} and 2MASS \citep[$70-1$\,$\mu$m;][]{For12}, RC3.9
  catalog \citep[extinction-corrected $V$, $B$, and $U$-bands;][]{deVaucouleurs91}, 
  ROSAT \citep[$0.1-2.4$\,keV;][]{Brinkmann94}, \textit{Suzaku}
  \citep[$2-10$\,keV;][]{Yang09}, \textit{Swift}-BAT
  \citep[$14-195$\,keV;][]{Baumgartner10} and LAT ($0.1-100$\,GeV;
  this paper). Gray curves correspond to the interstellar radiation 
  field/\texttt{GALPROP} models for the Milky Way \citep{Porter08,Str10} 
  placed at the distance of Circinus. Red curves
  denote the modeled emission of the accretion disk and the disk
  coronae using the \texttt{MyTORUS} model. Blue curves represent the modeled
  synchrotron and IC emission of the lobes and plumes. 
  The thick gray dotted line represents the $\gamma$-ray spectrum of 
  a starforming system (assumed $\Gamma = 2.2$) 
  corresponding to the IR luminosity of Circinus and IR-$\gamma$ 
  correlation found in \citet{Starforming}.
  \label{fig:SED}}
\end{center}
\end{figure}

Figure~\ref{fig:SED} presents the broad-band, multi-component SED of Circinus, 
including new LAT measurements. Integrated far-IR--to--optical measurements
represent the dominant starlight and the dust emission of the galaxy, 
with a negligible contribution from a heavily obscured AGN. In the X-ray regime, 
the hard X-ray fluxes are dominated by the heavily absorbed emission of the accretion 
disk and disk coronae. As demonstrated in \citet{Elmouttie98}, the total radio fluxes of 
the source in the $0.4-8.6$\,GHz range are due to a superposition 
of various emission components characterized by different spectral properties. 
Using their published ATCA 1.4, 2.4, 4.8, and 8.6\,GHz maps, 
we measured fluxes for each distinct component separately, 
namely for the nucleus (circle with radius, $r =1'' \simeq 20$\,pc), 
galaxy core including the central starburst region ($r=35''$ circle), 
the outer parts of the galaxy disk, NW lobe ($r=44.6''$ circle), 
NW plume ($50'' \times 25''$ box centered $68.15''$ from the nucleus), 
SE lobe (ellipse with radii of $70''$ and $55''$), 
and SE plume ($62.5'' \times 25''$ box centered $89.30''$ from the nucleus). 
The total radio emission of the Circinus system at $>$\,GHz frequencies 
is dominated by the central starburst region, 
and at lower frequencies by the outer parts of the galaxy disk; 
lobes and plumes contribute to the observed emission at the level of $10\%$; 
radio emission of the unresolved nucleus is negligible.

We fit the radio spectra of the lobes and plumes, assuming energy 
equipartition between the radiating electrons and the magnetic field (energy 
density ratio $u_e/u_B \equiv 1$), and a standard form of the electron 
energy distribution consisting of a power-law $dN_e/d\gamma \propto \gamma^{-2}$ 
between electron energies $\gamma_{\rm min} \equiv 1$ and $\gamma_{\rm br}$, 
breaking to $dN_e/d\gamma \propto \gamma^{-3}$ between $\gamma_{\rm br}$ 
and $\gamma_{\rm max} \equiv 10^6$. We adjust both the electron normalization 
and break Lorentz factors $\gamma_{\rm br}$ to obtain 
satisfactory fits to the radio data for each region separately, and then evaluate 
the expected IC emission. The results of the modeling are given as blue curves
in Figure~\ref{fig:SED}, where we show in addition the modeled emission of the accretion 
disk and the disk coronae using the \texttt{MyTORUS} model (red curves)~\citep{MyTorus}. For comparison, the figure
presents also the interstellar radiation field/\texttt{GALPROP} models \citep{Porter08,Str10} 
for the Milky Way placed at the distance of Circinus (gray curves). Note that the 
``clump'' structure seen around $10^{23}$\,Hz in the $\gamma$-ray continuum evaluated with 
\texttt{GALPROP} is due to the pion decay component dominating over the leptonic 
(IC and bremsstrahlung) ISM emission.	

The evaluated IC emission of the lobes severely underestimates the 
measured GeV emission of Circinus. In the model, magnetic field intensities read 
as $B \simeq 5-10$\,$\mu$G, 
in agreement with the values claimed by \citet{Elmouttie98}. The total energy stored in 
the entire structure is $E_{\rm tot} \simeq 10^{54}$\,erg. This is 
about an order of magnitude lower than the total energy of the lobes inferred by \citet{Mingo12} 
based on X-ray observations, indicating either a significant departure from the energy 
equipartition condition or dominant pressure support within the lobes provided 
by hot thermal plasma or relativistic protons. The departures from the minimum 
energy condition $u_e/u_B \gg 1$, which may enhance the expected IC radiation 
for a given synchrotron (radio) flux, are often claimed for lobes in radio galaxies 
and quasars \citep[see][and references therein]{Takeuchi12}. In the particular case of 
Circinus, however, the effect would have to be extreme in order to account 
for the flux detected with the LAT, $u_e/u_B \simeq 10^4$. Although we
cannot exclude this possibility, we consider it rather unlikely.  
Deep, high-resolution X-ray observations could in principle be used 
in the near future to validate the $u_e/u_B \gg 1$ hypothesis for the Circinus 
lobes, but at this moment the very limited photon statistics of the available 
\textit{Chandra} maps precludes any robust detection of a non-thermal
lobe-related emission component at keV photon energies. 
We note however that the IC model curve calculated for $u_e/u_B =1$ and 
shown in Figure~\ref{fig:SED} is below the corresponding upper limits.

Yet another possibility is the presence of an energetically 
significant population of CR protons associated with the radio
lobes. Relativistic protons injected by the kpc-scale jets/plumes, or
accelerated at the bow-shocks of the expanding lobes
\citep[see][]{Mingo12}, may generate a non-negligible $\gamma$-ray
emission due to the decay of pions produced during CR interactions
with surrounding matter. However, due to the sparseness of thermal gas
within the galactic halo (gas number density $\lesssim 10^{-3}$\,cm$^{-3}$), 
implying a low efficiency for this process, the resulting
GeV fluxes are likely below the $\gamma$-ray output of the galactic disk.

It is interesting to comment in this context on the similarities 
between the Circinus lobes and the \textit{Fermi}-LAT discovered giant `Fermi Bubbles' 
in our Galaxy \citep{dob10,su10}. Even though the origin of both structures is still under the debate, 
the jet activity of their central supermassive black holes --- which are of comparable masses, 
namely $\mathcal{M}_{\rm BH} \simeq (1.7 \pm 0.3) \times 10^6 \, M_{\odot}$ for Circinus \citep{Greenhill03} 
and $(4.5 \pm 0.4) \times 10^6 \, M_{\odot}$ for the Milky Way \citep[e.g.,][]{Ghez08} --- is the widely considered scenario. 
In the case of Circinus, the jet scenario is evidenced directly by the presence of 
collimated outflows supplying the lobes with energetic magnetized plasma 
\citep[see the discussion in][]{Elmouttie98,Mingo12}, while in the case of our Galaxy, 
it is supported by general energetic arguments and numerical simulations reproducing 
well the observed properties of bubbles \citep[e.g.,][]{Guo12,Yang12}. In both systems, 
the lobes extend to kpc scales across the galactic disks and are characterized by magnetic 
field strengths of the order of $\sim 10$\,$\mu$G \citep[see][for Fermi Bubbles]{su10,mer11}. 
Thus it seems that structures analogous to the Fermi Bubbles and Circinus lobes may not be uncommon 
in late-type galaxies undergoing episodic outbursts of AGN (jet) activity, but their 
contributions to the total $\gamma$-ray outputs of the systems do not exceed $\sim 10\%$.

\section{Conclusions}
\label{sec:concl}

Here we report the detection of a steady and spatially 
unresolved $\gamma$-ray source at the position of Circinus, 
consistent with 2FGL\,J1415.7$-$6520 based on a refined analysis 
using 4 years of LAT data.
Although the observed power-law spectrum ($\Gamma = 2.19\pm0.12$) is
similar to that of other LAT-detected starburst systems, Circinus is
$\gamma$-ray over-luminous by a factor of 5--6 relative to what is 
expected for emission from the ISM, based on multiwavelength correlations 
observed for nearby star-forming galaxies. However, the range of
SFRs estimated from radio, mid-IR, and far-IR luminosities span
a factor of $\simeq 4$, indicating large uncertainties in the expected CR
power delivered to the ISM.
We presented several alternative possibilities for the origin of the
GeV excess, including emission from the extended radio lobes, but
found no conclusive answer to fully account for the
GeV emission. This issue may be resolved by future studies, as Circinus is a 
compelling target for observations in the very high energy $\gamma$-ray 
regime with ground-based Cherenkov telescopes; the TeV-detected 
starburst systems NGC\,253 and M82 \citep{HESS-253,VERITAS-M82}
are actually fainter in the GeV range.

\section*{Acknowledgments}

\L.~S. was supported by Polish NSC grant DEC-2012/04/A/ST9/00083.
Work by C.C.C. at NRL is supported in part by NASA DPR S-15633-Y.

The \textit{Fermi} LAT Collaboration acknowledges generous ongoing support
from a number of agencies and institutes that have supported both the
development and the operation of the LAT as well as scientific data analysis.
These include the National Aeronautics and Space Administration and the
Department of Energy in the United States, the Commissariat \`a l'Energie Atomique
and the Centre National de la Recherche Scientifique / Institut National de Physique
Nucl\'eaire et de Physique des Particules in France, the Agenzia Spaziale Italiana
and the Istituto Nazionale di Fisica Nucleare in Italy, the Ministry of Education,
Culture, Sports, Science and Technology (MEXT), High Energy Accelerator Research
Organization (KEK) and Japan Aerospace Exploration Agency (JAXA) in Japan, and
the K.~A.~Wallenberg Foundation, the Swedish Research Council and the
Swedish National Space Board in Sweden.

Additional support for science analysis during the operations phase is gratefully
acknowledged from the Istituto Nazionale di Astrofisica in Italy and the Centre National d'\'Etudes Spatiales in France.

We thank M. Elmouttie for providing the ATCA images.

{}

\end{document}